\def\year{2022}\relax
   \documentclass[letterpaper]{article}
   \usepackage{aaai22} 
   \usepackage{times} 
   \usepackage{helvet} 
   \usepackage{courier} 
   \usepackage[hyphens]{url} 
   \usepackage{graphicx} 
   \usepackage{graphicx}  
   \usepackage{natbib}  
   \usepackage{caption}  
   \DeclareCaptionStyle{ruled}%
{labelfont=normalfont,labelsep=colon,strut=off} 
\usepackage{tikz}
\usepackage{cite}
\usepackage{soul}
\usepackage{natbib} 
\usepackage{xcolor}
\usepackage{subfig}
\usepackage{amsmath}
\usepackage{physics}
\usepackage{booktabs}
\usepackage{amsfonts}
\usepackage{textcomp}
\usepackage{algorithmic}
\usepackage{filecontents}
\usepackage[linesnumbered,lined,boxed,commentsnumbered]{algorithm2e}
\DeclareCaptionStyle{ruled}{labelfont=normalfont,labelsep=colon,strut=off} 
\begin{document}

\title{QUILT: Effective Multi-Class Classification on Quantum Computers Using an Ensemble of Diverse Quantum Classifiers}

\pdfinfo{
/Title (QUILT: Effective Multi-Class Classification on Quantum Computers Using an Ensemble of Diverse Quantum Classifiers)
/Author (Daniel Silver, Tirthak Patel, Devesh Tiwari) /TemplateVersion (2022.1)
}

\author {
    Daniel Silver,
    Tirthak Patel,
    Devesh Tiwari
}
\affiliations {
  Northeastern University\\
  \{silver.da, patel.ti, d.tiwari\}@northeastern.edu
}

\maketitle

\begin{abstract}
Quantum computers can theoretically have significant acceleration over classical computers; but, the near-future era of quantum computing is limited due to small number of qubits that are also error prone. Quilt is a framework for performing multi-class classification task designed to work effectively on current error-prone quantum computers. Quilt is evaluated with real quantum machines as well as with projected noise levels as quantum machines become more noise-free. Quilt demonstrates up to 85\% multi-class classification accuracy with the MNIST dataset on a five-qubit system.

\end{abstract}

\section{Introduction}
\label{sec:intro}

\textbf{Problem Motivation.} Quantum algorithms have been developed in the areas of chemistry, cryptography, machine learning, and optimization~\cite{Lu_Liu_Wang_Huang_Lin_He_2019,shor,Tiwari_Melucci_2019,Khairy_Shaydulin_Cincio_Alexeev_Balaprakash_2020}. A class of algorithms known as quantum variational algorithms are designed to optimize and execute quantum machine learning and classification workloads~\cite{qvc_paper}. 

While theoretically-promising, existing quantum machine learning classifiers are designed for future large-scale ideal quantum systems. This is because loading data, training, and testing samples on existing near-term intermediate-scale quantum (NISQ) computers is challenging due to significant hardware errors~\cite{schuld2019quantum,Jurcevic_2021,Preskill2018quantumcomputingin}. \emph{Consequently, existing quantum classifiers have been demonstrated to be effective only for relatively simple binary classification tasks~\cite{Schuld_2017,nature}. As our evaluation confirms, existing state-of-the-art approach are ineffective for multi-class classification (e.g., $<$30\% accuracy for eight-class image classification). Currently, there is lack of capability to perform multi-class classification tasks on the real quantum machines for exploration and improvement.}\vspace{1mm}

\noindent\textbf{Contributions.} Quilt specifically bridges this gap by open-sourcing its framework and dataset to the community for multi-class classification on NISQ quantum machines. Quilt makes the following key contributions:\vspace{1mm} 

(1) A key idea behind Quilt is to build an ensemble of quantum classifiers to perform multi-class classification. While intuitive and widely-applied in the classical world, we found that classical way of building ensembles is ineffective -- it improves the accuracy only by 5\%-10\% over the state of the art. To mitigate this challenge, Quilt builds a novel ensemble of quantum classifiers by (1) constructing diverse classifiers, and (2) keeping them simple.

(2) Second, we demonstrate that Quilt further enhances the classification accuracy obtained via carefully constructed diverse ensembles of models. Quilt makes an observation that when outputs from diverse models are combined, individual models agree on ``most parts'' of the output, but they may disagree on ``some parts'' due to the intrinsic noise, which cannot be mitigated simply by increasing the diversity or number of members. To mitigate this, Quilt designs a safeguard error correcting mechanism that detects which parts have most disagreement and then, builds specific error-correcting binary classifiers to increase the consensus for those parts, improving the overall prediction accuracy. 

(3) Third, we show how Quilt performs on a variety of datasets and in different simulation environments. We evaluate Quilt on MNIST, Fashion-MNIST, and Cifar for four and 8 classes. Quilt demonstrates an advantage in every case by up to 46\% over state-of-the-art for eight classes of MNIST.  We also demonstrate that Quilt can be run on real quantum machines today and test on IBM Manila and IBM Lima.

\section{Quantum Classification Background}

\textbf{Qubit and State Manipulation.} The qubit is the fundamental building block of a quantum computer. Where a classical bit is strictly binary, a qubit can be in a \textit{superposition} of both binary states, as shown below in Dirac notation:
\begin{equation}
\ket{\Psi} = \alpha\ket{0} + \beta\ket{1} \ s.t \ \ \alpha, \beta \in \mathbb{C} \ \& \ \ \norm{\alpha}^2 + \norm{\beta}^2 = 1
\end{equation}
Here, $\ket{\Psi}$ is the state of the qubit, which is in a superposition of the $\ket{0}$ and $\ket{1}$ basis states. When the qubit's state is measured, its superposition collapses and outputs state $\ket{0}$ with probability $\norm{\alpha}^2$ and $\ket{1}$ with probability $\norm{\beta}^2$.

Qubit states can be manipulated using unitary transformations that act as single- or multi-qubit quantum gates. Single-qubit gates are represented as rotation gates with three angle parameters: $R_3(\theta_1,\theta_2,\theta_3)$. Arbitrary superpositions can be achieved by tuning the rotation angles of this $R_3$ gate. Gates spanning multiple qubits have the ability to \textit{entangle} qubits into a unified quantum system.\vspace{1mm} 

\noindent\textbf{Quantum Circuits.} A quantum circuit is a sequence of gates applied to a system of qubits to affect the outcome of their entangled superposition state. The goal of this paper is to perform classification. On quantum computing systems, a specific type of quantum circuits -- quantum variational circuits (QVCs) -- have been used for classification tasks in place of classical neural networks due to their more efficient structure using far fewer parameters and their theorized exponential scaling properties over classical algorithms\cite{chen2020variational,Lockwood_Si_2020}. Therefore, next we briefly introduce the basic functions of QVCs. \vspace{1mm} 

\begin{figure}[t]
    \centering
    \includegraphics[scale=.474]{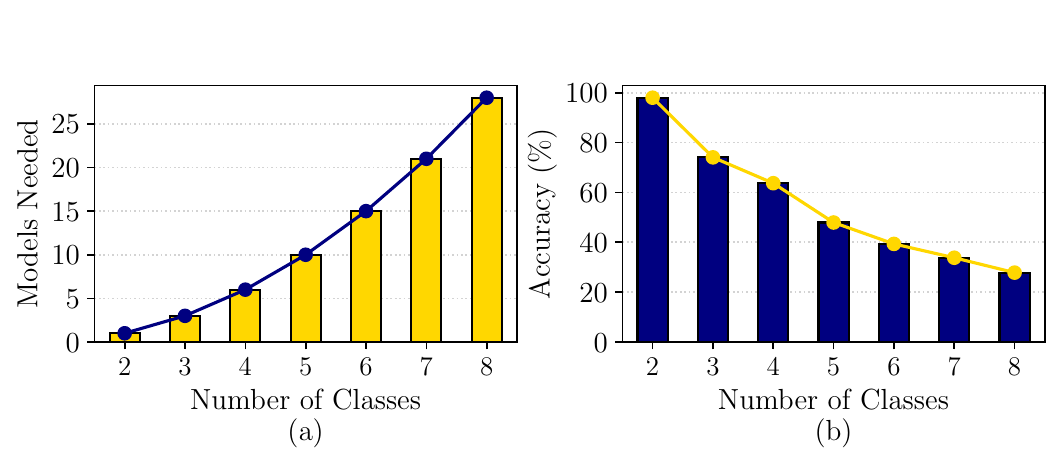}
    \caption{State-of-the-art multi-class quantum classifiers have (a) quadratic scaling for number of models and (b) quickly degrading accuracy scaling with number of classes.}
    \label{fig:onevone}
\end{figure}

\noindent\textbf{Quantum Variational Circuits.} \textit{QVCs are tunable quantum circuits that can be optimized for a specific purpose through training~\cite{Schuld_2019}.} QVCs are built using a mix of fixed gates (e.g., single-qubit gates with fixed rotation angles), which encode fixed classical data $x$ (e.g., input features used for a classification task) and tunable gates (e.g., single-qubit gates with parameterized rotation angles), which encode objective function information $\theta$ (e.g., maximize classification accuracy). A function $f$ is then created to map these parameters $(x,\theta)$ to a quantum system. This function can be any series of unitary operations that define the unique circuit architecture, applied to the initial $\ket{0}$ state.
\begin{equation}
f(x,\theta) = U(x,\theta)\ket{0}
\label{eq:2}
\end{equation}
The parameters of this function are tuned via a hybrid classical-quantum computing approach for objective optimization~\cite{qvc_paper,Havl_ek_2019}. QVCs are fully differentiable and can therefore, be optimized through standard deep learning optimization algorithms.

\noindent\textbf{QVCs for Classification.} QVCs are a suitable choice for a binary classifier as they provide high accuracy with few parameters, can be optimized with the same procedures as a neural network, and are small enough to fit on quantum machines today. A major challenge however is how to scale these classifiers to multi-class classification. Most real-world classification work done today involves more than two classes, but existing quantum classifiers are focused on solving only binary classification tasks and do not place much emphasis on multi-class classification tasks ~\cite{Schuld_2017} and~\cite{nature}.

\section{Motivation and Related Works}
\label{sec:motiv}
Quantum computing can have exponential acceleration over classical computers in many problem domains~\cite{shor,3body,many-body, qml,PhysRevLett.83.5162,Lu_Liu_Wang_Huang_Lin_He_2019}. Notably, quantum computing is also expected to prove useful for the field of machine learning and classification \cite{qml, fine_print,Lockwood_Si_2020, chen2020variational,Yang_Jiang_Zhang_Sun_2020,Li_Gkoumas_Sordoni_Nie_Melucci_2021,Khairy_Shaydulin_Cincio_Alexeev_Balaprakash_2020}. While new discoveries have led to better quantum machines, a defining limitation of the near-term noisy intermediate-scale quantum (NISQ) computers is the lack of large fault-tolerant  machines. One is therefore limited to using only a handful of qubits to perform quantum algorithms. This largely constrains the development of quantum algorithms for real-world applications. 

\begin{figure*}
\centering
  \includegraphics[scale=0.23]{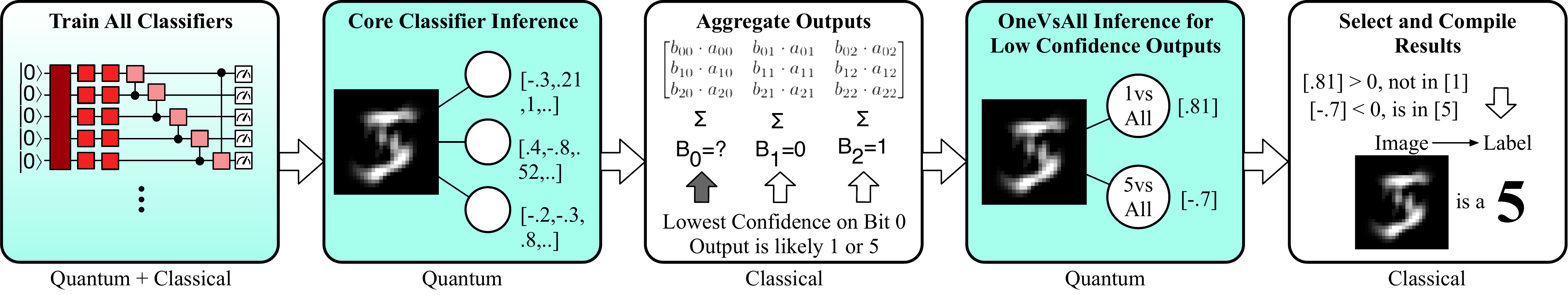}
  \caption{Overview of the design of Quilt. The label under each step indicates if the step is performed quantumly or classically.}
  \label{flow}
\end{figure*}

\subsubsection{Related Quantum Classification Efforts}
Multiple works have been put forward in the space of quantum image classification. For example, \cite{gambs2008quantum}, \cite{PhysRevLett.118.190503}, and \cite{wang2021quantum} are focused on theoretical algorithms that can only run on large fault-tolerant quantum computers; while promising, these works are not applicable in the current NISQ era as the computers are small and noisy. Other works leverage classical neural networks to handle much of the heavy lifting for quantum classification~\cite{8638596,9425825}. Unfortunately, in these cases, the potential for quantum speedup is greatly reduced as the works inherit the limitations of classical networks, such as when dealing with data in quantum entangled states \cite{nature}.

The most relevant and recent efforts in the quantum classification space are~\cite{Schuld_2017} and~\cite{nature}. These efforts are state-of-the-art effort for classification tasks on NISQ quantum computers. In \cite{Schuld_2017}, the authors use a simple distance-based classifier in simulation to classify the Iris dataset by analyzing how it performs for each pair of classes (1v2, 2v3, and 1v3) in separate binary classifications. The more recent work -- \cite{nature} -- tackles the same image classification problem of splitting the Iris dataset into tuples, with the addition of analyzing binary splits in the MNIST dataset. They use tensor networks to build their binary classifiers, then evaluate the performance on a real quantum computer. However, both~\cite{Schuld_2017} and~\cite{nature} are primarily designed to be effective for binary classification. While they attempt to extend to multi-class classification using binary classifications on every unique two-class pairing in their datasets, this ad-hoc extension approach has several limitations, as we discuss next and our evaluation confirms.  Recently, \cite{wang2021quantumnas} has proposed a framework to address noise in classification circuits.

\subsection{Why are the State-of-the-Art Quantum Classifiers Not Effective for Multi-Class Classification?}

The state-of-the-art approach for multi-class classification~\cite{nature} essentially relies on splitting a multi-class dataset into many binary datasets and classifying each pair and combining them all for a multi-class decision -- known as the OneVsOne Classifier. For example, for a three-class classification with classes A, B, and C, this would require three unique classifiers; for (A,B), (B,C), and (A,C). Each one votes in favor of one class and the class with two votes wins. However, this method has several shortcomings.

First, \textit{the current approach suffers from poor scalability as a large number of models are needed for training.} The number of models required to build an $n$-class multi-class classifier is $n(n-1)/2$ (Fig.~\ref{fig:onevone}(a)). Each model that needs to be trained classically slows down training significantly. Additionally, the inference on quantum machines becomes impractical with larger number of classes. For example, 8 classes would require 28 models, and 16 classes would require 120 different models to be initialized on quantum computers. Instead, \textit{Quilt, as we discuss later, is designed to scale only linearly with the number of classes}. %

Second, \textit{the accuracy of the current approach degrades quickly with increase in the number of classes (Fig.~\ref{fig:onevone}(b)).} One reason for this is the well-documented issue for the OneVsOne classifier: the ``the unclassable region'' problem~\cite{galar2011overview, liu2007nesting}. This is where no obvious winner can be selected from the binary classification pairs; for example, a three-class scenario where two candidates tie with one vote each. As more classes are added, the opportunities for ties increase as well. In fact, each additional classifier brings more noise to the system and adds more points of failures if the classifier is biased to a particular class. For this reason, \textit{we design Quilt with an ensemble of classifiers, where each classifier is wholly responsible to classify all classes}. 

\section{Quilt: Design and Implementation}
\label{sec:solut}

\textbf{Overview of the approach.} Fig.~\ref{flow} illustrates the overview of Quilt. A key idea behind Quilt is to use an ensemble of smaller classifier models instead of using a single large classifier model. This design decision is made to ensure that the impact of quantum operation errors -- which are prevalent on quantum real machines -- is minimized. Multiple lightweight simple models (referred to as ``core classifiers'') make the training and inference process simpler and reduce the impact of errors. In fact, at present, training and inferring large complex models is not feasible due to technological constraints and that is a major roadblock that Quilt curbs. 

A small-sized ensemble model is potentially less powerful than a complex, heavyweight model and may produce lower accuracy results. Therefore, to improve the classification accuracy, Quilt builds an ensemble in a unique way. As shown in the second stage of Fig.~\ref{flow}, each classifier outputs its classification prediction (three bits for an 8-class classification problem) and these inference results are combined across all classifiers by weighting according to their accuracies. As an efficiency measure, the already calculated outputs, $\hat{x}$, are used to calculate the accuracy after each iteration. We started with a withheld validation set (10\% of the data), but we robustly verified that it did not improve accuracy significantly; so we removed this calculation to save time. As we discuss in detail later, simply replicating the models is not sufficient for improving the accuracy. Quilt specifically designs them to be ``different by construction''. Each core classifier can have multiple instances and variants that help diversify the ensemble. Model, variant, and instance design details are discussed later in this section. 

During the development of Quilt, we discovered that building a diverse ensemble is also not sufficient. As the fourth stage of Fig.~\ref{flow} illustrates, when predictions across all classifiers are combined, some bits in the final predicted (e.g., one bit out of three bits for an 8-class prediction task) output may have low-confidence. This is despite careful selection and construction of individual core classifiers. To address this, Quilt designs unique binary classifiers to improve the prediction confidence of low-confidence bits. For example, in Fig.~\ref{flow}, ``1 vs All'' and ``5 vs All'' are used because the most significant bit has the lowest confidence and hence, the only output possibility is a `1' or a `5' depending upon the other predicted bits. Finally, Quilt combines the results to predict the target class in a multi-class setting.

Next, before we discuss the model design details of Quilt, we discuss the data processing procedure.


\begin{figure*}
    \centering
    \includegraphics[scale=0.35]{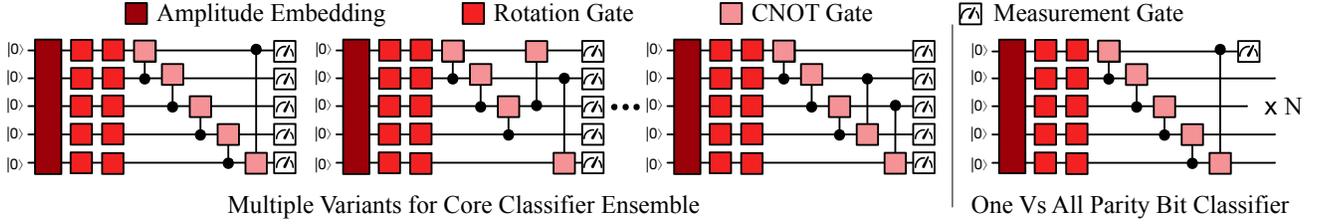}
    \caption{General structure of the different circuit architectures of Quilt.} 
    \label{circuit_diagram}
\end{figure*}
\subsection{Preprocessing and Embedding Images and Labels}
Running a quantum classifier on a small NISQ quantum machine requires loading and encoding classical data on the quantum computer in a scalable manner. Quilt has selected the model size to be five qubits because models spanning a larger number of qubits increase the surface area of error and hence, risk lowering the overall accuracy. 

\textit{Quilt's solution to encoding each image on a quantum machine is to perform Principal Component Analysis (PCA) on the entire dataset as a preprocessing procedure prior to training the model.} PCA allows particular specification for the exact number of dimensions for data reduction. PCA has a one-time offline overhead of processing the image data and generating the 32 features ($<1$ minute overhead even with serial processing of 60,000 MNIST images). This allows QUILT to fit on today’s NISQ machines and is the most efficient way to encode data to maximize explained variance across a dataset. It is worth noting that other works such as \cite{henderson2020quanvolutional} have used threshold based encoding where pixel values above a certain threshold are encoded as the state $\ket{0}$ and those above are in the state  $\ket{1}$. While it is possible this may perform well for certain datasets such as MNIST where color and shade are not as important, this should not be expected to perform well for more complex datasets where color is important.

Once the images are preprocessed, Quilt uses amplitude embedding to encode the normalized input vectors on to the qubits. Amplitude embedding transforms $L$ vectors of normalized classical data of size $N$ and encodes the data as a quantum state of size $log_2({NL})$.  The training set size affects the pre-processing step, but not the characteristics of individual quantum classifiers The first step of preparing classical data for quantum computation is to initialize all qubits to the ground state: $\ket{0}$.  Then, the state preparation circuit ($U_x$) is applied to add the fixed classical data to the circuit, embedding one input feature per quantum state amplitude. Once the state is prepared, Quilt's classification model can then be applied. 
Quilt also preprocesses the labels to ensure compatibility with the quantum space. This requires selecting a value of \{-1,1\} for each bit representing the label. With the preprocessing step complete, we delve into the model design details of Quilt.





\subsection{Quilt's Model Ensembles}



Large and complex models can be used as a building block for classification, but they increase the likelihood of errors on error-prone NISQ machines. \textit{To mitigate this challenge, Quilt builds on a base of small-sized circuits because smaller circuit size reduces the adverse impact of errors. In addition, to compensate for the small circuit size, Quilt builds an ensemble of such models that work together to collectively improve the classification performance.} Quilt uses accuracy as a weighting mechanism to aggregate the classifications of individual ensemble components. Each ensemble $\epsilon$ with weight $w_\epsilon$ contributes to the output $\hat{y}$ according to the following equation:
\begin{equation}
\hat{y} = sgn\Big[\sum_{\epsilon} w_\epsilon O(x, \theta)\Big]
\end{equation}
where the $sgn(z)$ operation is defined as 
\[
  sgn(z) = \left.
  \begin{cases}
    -1, & \text{for } z < 0 \\
     1, & \text{for } z > 0 
  \end{cases}
  \right\}
\]
and the output $O(x, \theta)$ is a real number between -1 and 1, with its absolute value suggesting the confidence of being a member of the -1 or 1 class. The output is the result of applying the $U(x,\theta)$ operator to the $\ket{0}$ state and measuring it, where $U$ is the unitary transformation applied to input $x$ and parameters $\theta$ (from Eq.~\ref{eq:2}).

The ensemble representation for a QVC then becomes
\begin{equation}
\hat{y} = sgn\Big[\sum_{\epsilon}\dfrac{a_\epsilon}{a}  O(x, \theta)\Big]
\label{eq:5}
\end{equation}
where $a_\epsilon$ is the associated accuracy of ensemble member $\epsilon$ and $a$ is the sum of all ensemble accuracies.
\subsection{Model Construction}
\noindent\textbf{Core Classifier.} The building block of Quilt's ensemble is the core classifier, which is designed to be able to classify all classes on its own (as opposed an ensemble of OneVsOne classifiers used for state-of-the-art multi-class classification). It consists of an amplitude embedding circuit, followed by repeating rotation gates on each qubit, then CNOT gates to entangle all qubits together, followed by a measurement on all output qubits. 

In classical computing, traditional neural networks perform classification by assigning a single class to an output neuron to then use a function such as Softmax to find the most likely candidate. \textit{Quilt uses a different approach by leveraging the quantum nature of the model: it uses each measurement qubit $b$ as a bit in the full $n$-bit classification. This allows better scaling to handle a larger number of classes.} All measurement qubits form a single bitstring that has an associated class label:
\begin{equation}
\hat{y} = \sum_{b=0}^{n}2^{n-b}\Big(\dfrac{sgn(q_b)+1}{2}\Big)
\end{equation}
Here, $q_b$ is the value of the $b^{th}$ bit as measured in the quantum state output (Eq.~\ref{eq:5}). The number of these models scale logarithmically with the number of classes in the label space $n=log_2{N}$, but the accuracy scales inversely. This is partially because in order for the entire output to be correct, all the bits have to be correct. For example, if each bit is correct with probability .95, the probability of a correct guess for a three-bit classifier would be $.95^3 = .857$ or 85.7\%.\vspace{1mm} 

\begin{figure}[t]
\centering
  \includegraphics[scale=0.205]{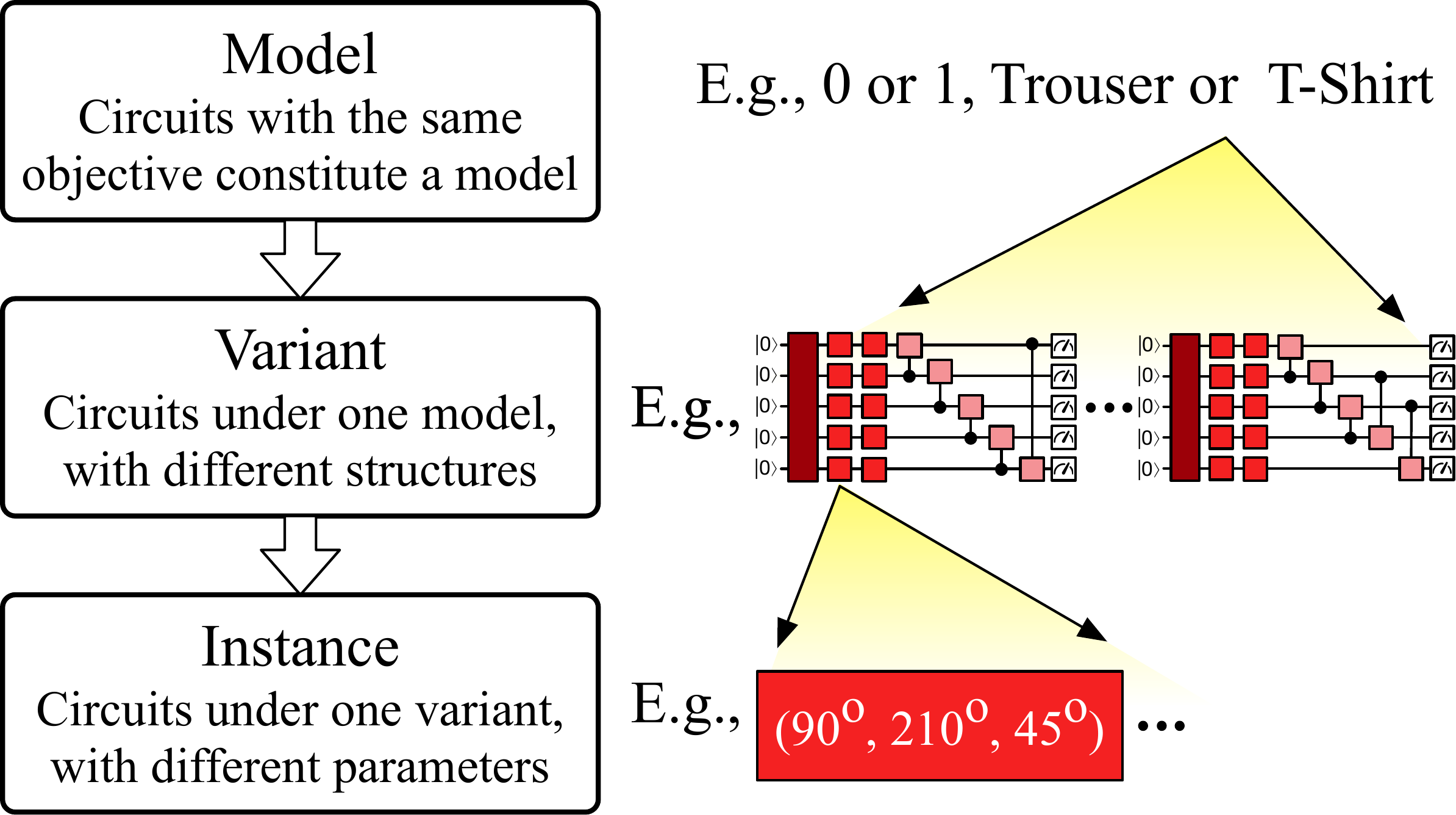}-
  \caption{Ensemble diversity hierarchy. At the top of the hierarchy is the ''model'', which represents a quantum circuit with a unique objective. An example of this would be a OneVsOne classifier with the objective of distinguishing Trousers and T-Shirts. A model can have multiple ''variants'', which are circuits with the same objective, but different structures. Finally, a variant can have multiple ''instances''. All instances associated with a variant have the same structure, but may differ in their parameters due to different initial conditions used for optimization.}
  \label{definitions}
\end{figure}

\noindent\textbf{Multiple Diverse Variants of Core Classifiers.} Now that we have the general structure for a core classifier defined, our next task is to use this general structure to form an ensemble. It is imperative to diversify the ensemble components as much as possible in order to ensure that they correct for each other's deficiencies, blind spots, and biases. Initializing model instances with different initial conditions enforces some degree of diversity. However, these instances would still be optimizing over the same search space due to the model structure being the same, thus resulting in similar outputs and reduced ensemble diversity.

Quilt mitigates this challenge by constructing model variants with heterogeneous architectures in order to establish more diversity and independence between them. These variants differ in that each variant under the same model can have different number of additional $CNOT$ gates and corresponding connections -- Fig. \ref{circuit_diagram} provides examples of different variants of the core classifier circuit. Each heterogeneous variant therefore has a different hyper-dimensional search space to optimize over and a different set of minima, resulting in a diverse set of core classifiers with different structures and parameters. The output of these variants can be combined by integrating Quilt's bitwise core classifiers with the weighted ensemble averaging, as expressed below:
\begin{equation}
\hat{y} = \sum_{b=0}^{n}2^{n-b}\Bigg[\dfrac{sgn\Big(\sum_{\epsilon}  \dfrac{a_\epsilon}{a}  q_{b\epsilon}\Big)+1}{2}\Bigg]
\end{equation}
Fig.~\ref{definitions} provides a visual representation of the diversity hierarchy of a model, a variant, and an instance. Different models are the most diverse due to them optimizing different objectives. However, for Quilt's ensemble, we need all core classifier models to have the same objective. Therefore, Quilt employs different variants that are part of the same model (i.e., have the same objective) and provide better diversity than just having different instances.\vspace{1mm}

\noindent\textbf{OneVsAll Classifiers for Error Correcting.} While core classifiers do most of the heavy lifting with classification, the addition of support networks improves the overall accuracy of the classification. \textit{One vs all (OneVsAll) classifiers are used for each class in order to provide a parity bit when necessary. These act as error correcting procedures that activate under certain conditions.} A OneVsAll classifier is a binary classifier where one class is assigned to one class and every other class is grouped together in the other class. In the case of Quilt, there is one OneVsAll classifier for each class in the class space. These classifiers only act when the core classifiers have low confidence, defined by a confidence threshold, $\gamma$. This is shown in Eq. \ref{gamma_eq} and is applied to each image separately.
\begin{equation}
\Big{|} \sum_{\epsilon}  \dfrac{a_\epsilon}{a}  O(x,\theta)\Big{|}  < \gamma
\label{gamma_eq}
\end{equation}
We find the value of $\gamma$ is optimal when it is met for 10\% of all guesses. For all images, if any have an absolute sum of weighted outputs that falls within $\gamma$, we identify the least confident bit $b$ in the output as:
\begin{equation}
b = \operatorname*{argmin}_{b} \Big{|}\sum_{\epsilon}  \dfrac{a_\epsilon}{a}  O(x,\theta)\Big{|}
\end{equation}
The next step is to evaluate the associated OneVsAll classifier for a final verdict on the bit in question. As there is a single bit in question, two OneVsAll Classifiers that are charged with the decision. For example, in the case of evaluating the most significant bit where the two least significant bits are certain (e.g., X01), the output is either a 1 or a 5, so the 1VsAll and 5VsAll classifiers must be evaluated. The bit is only set if and only if both OneVsAll Classifiers being evaluated agree on a decision, otherwise it defaults to its original state. This is shown for classifiers $C_1$ and $C_2$ below.
\[
  q_b = \left.
  \begin{cases}
     0, & \text{for } C_1 = -1, C_2 = 1 \\
     1, & \text{for } C_1 =  1, C_2 = -1 \\
     q_b, & \text{otherwise }
  \end{cases}
  \right\}
\]
Here, $-1$ indicates that the image is in the specific OneVsAll class, and $1$ indicates that it is not in the class.

\subsection{Stitching Together the Ensemble}

Overall, Quilt requires five core classifiers + $n$ OneVsAll classifiers to build an ensemble for $n$ classes. For example, for its eight-class ensemble, Quilt requires 13 classifiers: five core classifiers (each is a different variant) and eight OneVsAll classifiers. \textit{The number of models required by Quilt therefore scales linearly with the number of classes, as opposed to quadratically with the state-of-the-art approach (Fig.~\ref{fig:onevone})}. Quilt's models construct a robust scalable ensemble with error correction built in. Next, we describe the implementation of our training and inference procedure.


 
\subsection{Training and Inference Procedure}


Quilt begins by constructing the core model variants, each set to a learning rate of .05, optimized using the Adam optimizer~\cite{kingma2014adam}.  This is followed by randomly initializing all variants' weights separately between 0 and 1. For 100 epochs, these variants sample 50 random images and corresponding labels from the training set split. All vectorized images are passed into the network for evaluation. Within the context of the hybrid quantum-classical framework, gradients are computed classically after generating the outputs of the circuit. The new gate parameters are then optimized by shifting along the gradient to estimate the eigenvalues, then repeating until the optimization has terminated. Quilt uses a form of a squared $L_2$ loss to perform gradient descent optimization. We find that squared loss works the best in training these small classification networks and outperforms $L_1$ loss or even techniques enforcing sparsity, such as the $L_0$ norm. Quilt incorporates $L_2$ loss in multiclass classification using $N$ outputs qubits as follows:
\begin{equation}
\text{Loss} = {\sum_{i=0}^{\text{batch}}\sum_{j=0}^{N}}{(\text{label}[i][j] -\text{prediction}[i][j])^2}
\end{equation}

As an instance, in the eight-class case, this translates to only training the optimizer against the outputs of the first three qubits and ignoring the last two. As training these types of sensitive quantum networks does not provide a reliable advantage from one epoch to the next, we save the weights for the running best model.

To train the OneVsAll classifiers, we sample the labels to create an equal amount of data in both classes of the classifier. For example, to classify a ``7'' with 1000 images, the training set would consist of 1000 images of ``7'' and 1000 images composed of all other digits. These classifiers are also optimized using the Adam optimizer with $L_2$ loss.

In terms of inference, Quilt runs the five trained core classifiers for each inference task and takes the weighted average of their output. If it determines that there exist bits with low confidence, Quilt runs the respective OneVsAll classifiers for the bit with the lowest confidence. Once the final output is produced, Quilt assesses all the classifier outputs and generates the final label for the inference task.
\section{Evaluation}
\label{sec:evalu}

\subsection{Experimental Methodology}

\noindent\textbf{Dataset.} We evaluate Quilt with the MNIST \cite{mnist}, Fashion-MNIST \cite{xiao2017/online}, and Cifar \cite{cifar} datasets, using an 80/20 training-to-testing split. MNIST is a greyscale digit dataset that is widely used for quantum classification \cite{nature,wilson2018quantum,PhysRevA.101.062327,Li_Song_Wang_2021,Tiwari_Melucci_2019}. We also use Fashion-MNIST and Cifar due to the greater difficult of classifying them than MNIST. For example, Cifar has more pixels than MNIST and is also in RGB. We use classes [\{(0, 1), 6, 7\}, 2, 3, 4, 5] for MNIST, [\{(t-shirt/top, trouser), pullover, dress\}, coat, sandal, shirt, sneaker] for Fashion-MNIST, and [\{(airplane, automobile), bird, cat\}, deer, dog, frog, horse] for Cifar. Here, the class sets used for two-class, four-class, and eight-class classification are enclosed in (), \{\}, and [], respectively. We refer to these classes using the naming convention ``$<$dataset$>$-$<$classes$>$'' (e.g., Cifar-8 for eight-class Cifar).\vspace{1mm}


\noindent\textbf{Setup Details.} Our solution environment uses Python3, with Pennylane \cite{pennylane} as a framework using Qiskit \cite{Qiskit} as the backbone. We use Sklearn to perform PCA preprocessing for our datasets. Our noise-free simulations take place using the default Pennylane backened simulator. For our noisy simulations, we use ``qiskit.aer'' to set the error rates of quantum gates. Our real machine evaluations are run on IBM's five-qubit Lima and Manila computers.\vspace{1mm}

\noindent\textbf{Comparative Techniques.} (1) \textbf{OneVsOne:} As introduced in Fig.~\ref{fig:onevone}, this current state-of-the-art technique performs multi-class classification by combining many binary classifiers for a multi-class decision. This is the extension of ~\cite{nature}. (2) \textbf{Ensemble:} The ensemble is similar to Quilt, but without the OneVsAll classifiers or multiple variants. Each member is responsible for a bit of the classification, allowing the binary classifier to scale to multi-class. There are five of these members for each bit and the decision procedure is a simple majority rules. \textit{We include this method for a comparison to show why off-the-shelf ensemble methods are not sufficient for multi-class classification.} We use the same PCA-based dataset preprocessing technique as Quilt for the competitive techniques for fair comparison. 

\subsection{Results and Analysis}

\begin{figure}[t]
    \centering
    \includegraphics[scale=.529]{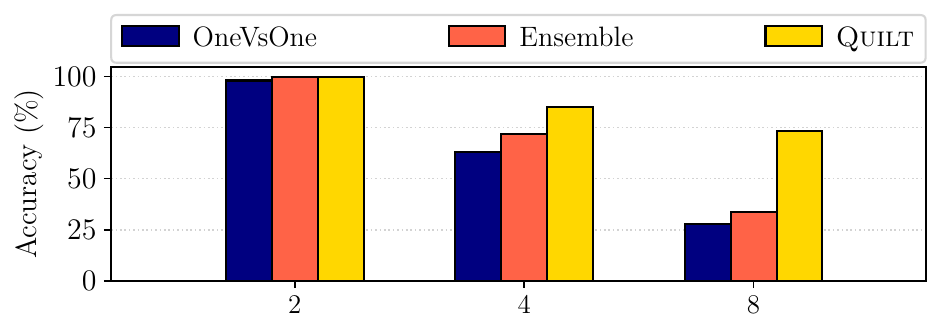}
    \caption{Quilt scales better to higher number of classes than comparative techniques for the MNIST dataset.}
    \label{mnistall}
\end{figure}

\noindent\textbf{Quilt achieves higher accuracy than comparative state-of-the-art techniques as the number of classes increases.} We show classification accuracy for MNIST-2, MNIST-4, and MNIST-8 in Fig. \ref{mnistall}. We make several observations. We note that all methods perform similarly on 2 classes as binary classification is a relatively simpler task. However, Quilt provides 22\% and 46\% point better classification accuracy over the state-of-the-art OneVsOne method as the number of classes increases to four and eight. This is because the accuracy-weighted ensemble allows Quilt to factor in model accuracy, allowing more accurate models to be favored. Additionally, Quilt consolidates classification to an ensemble, where each model is responsible for classifying all classes on its own, therefore making better use of the full capacity of the QVC: \textit{where Quilt only requires 13 circuits for eight-class classification, the state-of-the-art OneVsOne classifier requires 28}. However, having an ensemble is not enough; by checking the error-prone bits, Quilt is able to effectively treat one of the bits in each bad sample and obtain higher accuracy than the ensemble method. Even though the simple ensemble approach requires fewer models (i.e., $5$), we find Quilt's linear increase in the number of models with the number of classes to be justified given the significant increase in accuracy.\vspace{1mm}


\begin{figure}[t]
    \centering
    \includegraphics[scale=.537]{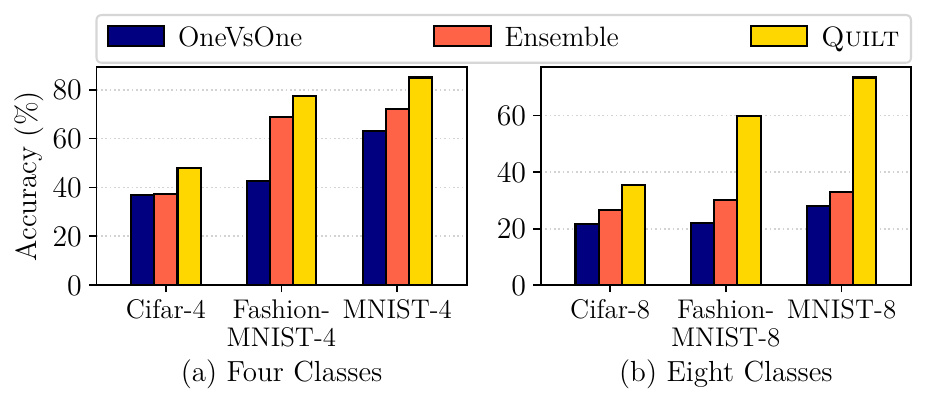}
    \caption{Quilt's accuracy is better than other methods for all well-known (a) four-class and (b) eight-class datasets.}
    \label{4digit}
\end{figure}

\begin{figure}[t]
    \centering
    \includegraphics[scale=.537]{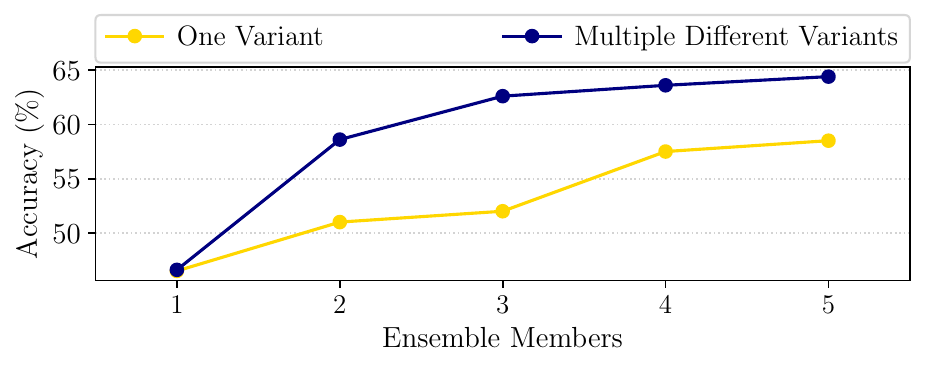}
    \caption{An ensemble with different variants outperforms one with multiple instances of one variant (MNIST-4).}
    \label{samevsdifferentarc}
\end{figure}

\begin{figure}[t]
    \centering
    \includegraphics[scale=.537]{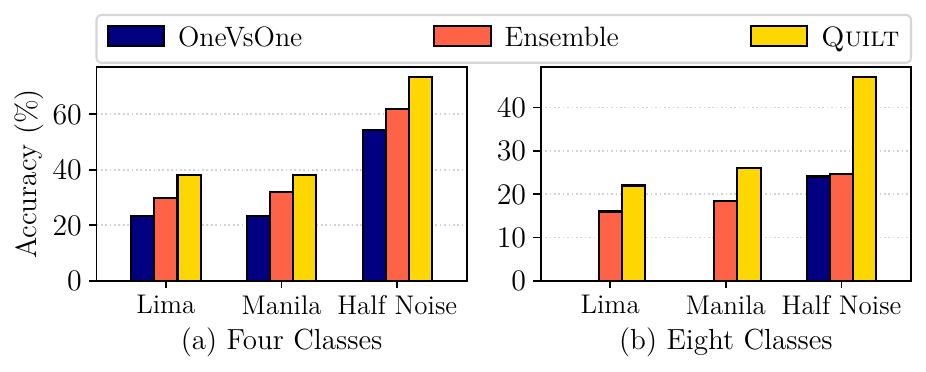}
    \caption{(a) MNIST-4 and (b) MNIST-8 accuracy on IBM's quantum machines Lima and Manila along an interpolation based on half of Manila's noise levels. OneVsOne could not be computed for eight classes on the real machines as it required an impractical number of model evaluations.}
    \label{real-4}
\end{figure}

\begin{figure}[t]
    \centering
    \includegraphics[scale=.537]{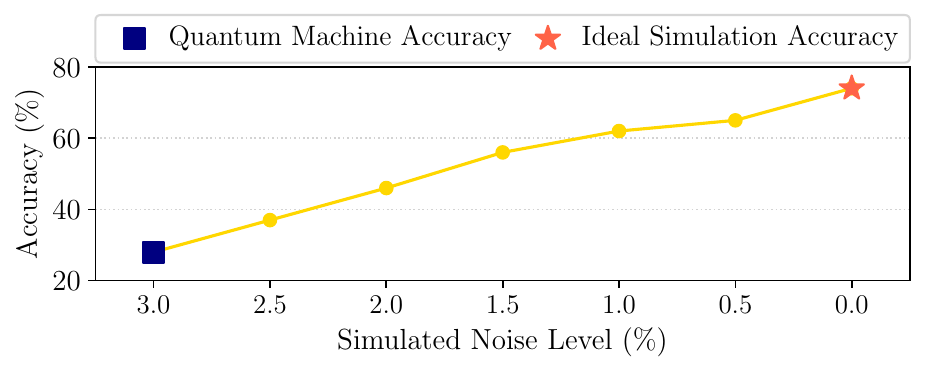}
    \caption{Projected increase in MNIST-8 accuracy as the hardware noise decreases with evolution in hardware.}
    \label{simulatednoise}
\end{figure}

\noindent\textbf{Quilt outperforms competitive techniques across multiple well-known deep learning image datasets.} In  Fig. \ref{4digit}(a), we evaluate how well Quilt performs on Fashion-MNIST-4 and Cifar-4, in addition to MNIST-4.  We also look at how Quilt compares to other solutions. Quilt outperforms the state-of-the-art by 46\%, 38\%, and 14\% for MNIST-8, Fashion-MNIST-8, and Cifar-8, respectively. All methods perform better on Fashion-MNIST-4 and MNIST-4 than on Cifar-4. This is because of the limited size of the encoded data that makes encoding the RGB images in Cifar-4 into 32 features tougher (i.e., more information is lost by performing PCA on Cifar images). We also evaluate performance of different datasets for eight classes in Fig. \ref{4digit}(b) and observe similar behavior. Comparing the two figures demonstrates that Quilt is the most scalable and highest performing solution across different datasets.\vspace{1mm}

\noindent\textbf{Quilt's strategy of using multiple variants achieves significantly higher accuracy than using a single variant.} Fig. \ref{samevsdifferentarc} confirms Quilt's critical design component: higher accuracy is achieved by adding diversity to variant architectures in ensembles. While different instances of a homogeneous variant can be diversified using different initial weights, because these instances traverse the same hyper-dimensional space during parameter optimization, they cannot provide the level of diversity of heterogeneous variants.\vspace{1mm}


\noindent\textbf{Quilt outperforms other techniques on running image classification tasks on today's noisy real quantum machines.} Fig. \ref{real-4} (a) and (b) show Quilt's classification accuracy for MNIST-4 and MNIST-8 datasets on real quantum computers. Quilt outperforms other methods due to the redundancies built into it that were designed to make it work well on small and noisy circuits. While we do not focus our optimizations on the noisy machines of today, we still abide by the constraints that allow Quilt to be executable on today's quantum machines. We also note that the overall accuracy on real systems in lower than our previous simulated results due to the hardware noise. However, in the future, the noise is expected to reduce significantly, with estimates such as $10^{-4}$ error rate ~\cite{Jurcevic_2021}. \vspace{1mm}

\noindent\textbf{Quilt's accuracy performance will only improve with reduction in hardware noise levels as quantum technology advances.} We simulate several noise levels between the ideal case of 0 and the current case of real machines and show how different levels of noise affect the accuracy. Fig. \ref{simulatednoise} demonstrates that as noise levels continue to improve on real NISQ devices, the accuracy of Quilt improves. 


\section{Conclusion}
\label{sec:concl}

In this paper, we proposed Quilt, an ensemble-based end-to-end multi-class image classifier for NISQ machines. We showed the scalability potential of Quilt compared to other methods and demonstrated a significant improvement over the state-of-the-art in terms of accuracy and resource requirements. We also showed that these models can be run on today’s quantum machines and project the improvements of QUILT as quantum machines become more error-resistant.

\section*{Acknowledgements}

We are thankful for the thoughtful feedback from our anonymous reviewers and the support from Northeastern University, NSF Award 1910601, and the Massachusetts Green High Performance Computing Center (MGHPCC) facility. We acknowledge the use of the IBM Q for this work. The views expressed are those of the authors and do not reflect the official policy or position of IBM or the IBM Q team.
\bibliography{aaai22}

\begin{thebibliography}{39}
\providecommand{\natexlab}[1]{#1}

\bibitem[{Aaronson(2015)}]{fine_print}
Aaronson, S. 2015.
\newblock Read the fine print.
\newblock \emph{Nature Physics}, 11(4): 291--293.

\bibitem[{Abraham et~al.(2019)Abraham, AduOffei, Agarwal, Akhalwaya, Aleksandrowicz, Alexander, Amy, Arbel, Arijit02, Asfaw, Avkhadiev, Azaustre, AzizNgoueya, Banerjee, Bansal, Barkoutsos, Barnawal, Barron, Barron, Bello, Ben-Haim, Bevenius, Bhobe, Bishop, Blank, Bolos, Bosch, Brandon, Bravyi, Bryce-Fuller, Bucher, Burov, Cabrera, Calpin, Capelluto, Carballo, Carrascal, Chen, Chen, Chen, Chen, Chen, Chow, Churchill, Claus, Clauss, Cocking, Correa, Cross, Cross, Cross, Cruz-Benito, Culver, C{\'o}rcoles-Gonzales, Dague, Dandachi, Daniels, Dartiailh, DavideFrr, Davila, Dekusar, Ding, Doi, Drechsler, Drew, Dumitrescu, Dumon, Duran, EL-Safty, Eastman, Eberle, Eendebak, Egger, Everitt, Fern{\'a}ndez, Ferrera, Fouilland, FranckChevallier, Frisch, Fuhrer, Fuller, GEORGE, Gacon, Gago, Gambella, Gambetta, Gammanpila, Garcia, Garg, Garion, Gilliam, Giridharan, Gomez-Mosquera, Gonzalo, de~la Puente~Gonz{\'a}lez, Gorzinski, Gould, Greenberg, Grinko, Guan, Gunnels, Haglund, Haide, Hamamura, Hamido, Harkins, Havlicek,
  Hellmers, Herok, Hillmich, Horii, Howington, Hu, Hu, Huang, Huisman, Imai, Imamichi, Ishizaki, Iten, Itoko, JamesSeaward, Javadi, Javadi-Abhari, Javed, Jessica, Jivrajani, Johns, Johnstun, Jonathan-Shoemaker, K, Kachmann, Kale, Kanazawa, Kang-Bae, Karazeev, Kassebaum, Kelso, King, Knabberjoe, Kobayashi, Kovyrshin, Krishnakumar, Krishnan, Krsulich, Kumkar, Kus, LaRose, Lacal, Lambert, Lapeyre, Latone, Lawrence, Lee, Li, Liu, Liu, Maeng, Majmudar, Malyshev, Manela, Marecek, Marques, Maslov, Mathews, Matsuo, McClure, McGarry, McKay, McPherson, Meesala, Metcalfe, Mevissen, Meyer, Mezzacapo, Midha, Minev, Mitchell, Moll, Montanez, Monteiro, Mooring, Morales, Moran, Motta, MrF, Murali, M{\"u}ggenburg, Nadlinger, Nakanishi, Nannicini, Nation, Navarro, Naveh, Neagle, Neuweiler, Nicander, Niroula, Norlen, NuoWenLei, O'Riordan, Ogunbayo, Ollitrault, Otaolea, Oud, Padilha, Paik, Pal, Pang, Pascuzzi, Perriello, Phan, Piro, Pistoia, Piveteau, Pocreau, Pozas-iKerstjens, Prokop, Prutyanov, Puzzuoli, P{\'e}rez, Quintiii,
  Rahman, Raja, Ramagiri, Rao, Raymond, Redondo, Reuter, Rice, Riedemann, Rocca, Rodr{\'\i}guez, RohithKarur, Rossmannek, Ryu, SAPV, SamFerracin, Sandberg, Sandesara, Sapra, Sargsyan, Sarkar, Sathaye, Schmitt, Schnabel, Schoenfeld, Scholten, Schoute, Schwarm, Sertage, Setia, Shammah, Shi, Silva, Simonetto, Singstock, Siraichi, Sitdikov, Sivarajah, Sletfjerding, Smolin, Soeken, Sokolov, Sokolov, SooluThomas, Starfish, Steenken, Stypulkoski, Sun, Sung, Takahashi, Takawale, Tavernelli, Taylor, Taylour, Thomas, Tillet, Tod, Tomasik, de~la Torre, Trabing, Treinish, TrishaPe, Tulsi, Turner, Vaknin, Valcarce, Varchon, Vazquez, Villar, Vogt-Lee, Vuillot, Weaver, Weidenfeller, Wieczorek, Wildstrom, Winston, Woehr, Woerner, Woo, Wood, Wood, Wood, Wood, Wootton, Yeralin, Yonge-Mallo, Young, Yu, Zachow, Zdanski, Zhang, Zoufal, Zoufalc, a~kapila, a~matsuo, bcamorrison, brandhsn, nick bronn, brosand, chlorophyll zz, csseifms, dekel.meirom, dekelmeirom, dekool, dime10, drholmie, dtrenev, ehchen, elfrocampeador,
  faisaldebouni, fanizzamarco, gabrieleagl, gadial, galeinston, georgios ts, gruu, hhorii, hykavitha, jagunther, jliu45, jscott2, kanejess, klinvill, krutik2966, kurarrr, lerongil, ma5x, merav aharoni, michelle4654, ordmoj, sagar pahwa, rmoyard, saswati qiskit, scottkelso, sethmerkel, shaashwat, sternparky, strickroman, sumitpuri, tigerjack, toural, tsura crisaldo, vvilpas, welien, willhbang, yang.luh, yotamvakninibm, and {\v{C}}epulkovskis}]{Qiskit}
Abraham, H.; AduOffei; Agarwal, R.; Akhalwaya, I.~Y.; Aleksandrowicz, G.; Alexander, T.; Amy, M.; Arbel, E.; Arijit02; Asfaw, A.; Avkhadiev, A.; Azaustre, C.; AzizNgoueya; Banerjee, A.; Bansal, A.; Barkoutsos, P.; Barnawal, A.; Barron, G.; Barron, G.~S.; Bello, L.; Ben-Haim, Y.; Bevenius, D.; Bhobe, A.; Bishop, L.~S.; Blank, C.; Bolos, S.; Bosch, S.; Brandon; Bravyi, S.; Bryce-Fuller; Bucher, D.; Burov, A.; Cabrera, F.; Calpin, P.; Capelluto, L.; Carballo, J.; Carrascal, G.; Chen, A.; Chen, C.-F.; Chen, E.; Chen, J.~C.; Chen, R.; Chow, J.~M.; Churchill, S.; Claus, C.; Clauss, C.; Cocking, R.; Correa, F.; Cross, A.~J.; Cross, A.~W.; Cross, S.; Cruz-Benito, J.; Culver, C.; C{\'o}rcoles-Gonzales, A.~D.; Dague, S.; Dandachi, T.~E.; Daniels, M.; Dartiailh, M.; DavideFrr; Davila, A.~R.; Dekusar, A.; Ding, D.; Doi, J.; Drechsler, E.; Drew; Dumitrescu, E.; Dumon, K.; Duran, I.; EL-Safty, K.; Eastman, E.; Eberle, G.; Eendebak, P.; Egger, D.; Everitt, M.; Fern{\'a}ndez, P.~M.; Ferrera, A.~H.; Fouilland, R.;
  FranckChevallier; Frisch, A.; Fuhrer, A.; Fuller, B.; GEORGE, M.; Gacon, J.; Gago, B.~G.; Gambella, C.; Gambetta, J.~M.; Gammanpila, A.; Garcia, L.; Garg, T.; Garion, S.; Gilliam, A.; Giridharan, A.; Gomez-Mosquera, J.; Gonzalo; de~la Puente~Gonz{\'a}lez, S.; Gorzinski, J.; Gould, I.; Greenberg, D.; Grinko, D.; Guan, W.; Gunnels, J.~A.; Haglund, M.; Haide, I.; Hamamura, I.; Hamido, O.~C.; Harkins, F.; Havlicek, V.; Hellmers, J.; Herok, {\L}.; Hillmich, S.; Horii, H.; Howington, C.; Hu, S.; Hu, W.; Huang, J.; Huisman, R.; Imai, H.; Imamichi, T.; Ishizaki, K.; Iten, R.; Itoko, T.; JamesSeaward; Javadi, A.; Javadi-Abhari, A.; Javed, W.; Jessica; Jivrajani, M.; Johns, K.; Johnstun, S.; Jonathan-Shoemaker; K, V.; Kachmann, T.; Kale, A.; Kanazawa, N.; Kang-Bae; Karazeev, A.; Kassebaum, P.; Kelso, J.; King, S.; Knabberjoe; Kobayashi, Y.; Kovyrshin, A.; Krishnakumar, R.; Krishnan, V.; Krsulich, K.; Kumkar, P.; Kus, G.; LaRose, R.; Lacal, E.; Lambert, R.; Lapeyre, J.; Latone, J.; Lawrence, S.; Lee, C.; Li, G.; Liu,
  D.; Liu, P.; Maeng, Y.; Majmudar, K.; Malyshev, A.; Manela, J.; Marecek, J.; Marques, M.; Maslov, D.; Mathews, D.; Matsuo, A.; McClure, D.~T.; McGarry, C.; McKay, D.; McPherson, D.; Meesala, S.; Metcalfe, T.; Mevissen, M.; Meyer, A.; Mezzacapo, A.; Midha, R.; Minev, Z.; Mitchell, A.; Moll, N.; Montanez, J.; Monteiro, G.; Mooring, M.~D.; Morales, R.; Moran, N.; Motta, M.; MrF; Murali, P.; M{\"u}ggenburg, J.; Nadlinger, D.; Nakanishi, K.; Nannicini, G.; Nation, P.; Navarro, E.; Naveh, Y.; Neagle, S.~W.; Neuweiler, P.; Nicander, J.; Niroula, P.; Norlen, H.; NuoWenLei; O'Riordan, L.~J.; Ogunbayo, O.; Ollitrault, P.; Otaolea, R.; Oud, S.; Padilha, D.; Paik, H.; Pal, S.; Pang, Y.; Pascuzzi, V.~R.; Perriello, S.; Phan, A.; Piro, F.; Pistoia, M.; Piveteau, C.; Pocreau, P.; Pozas-iKerstjens, A.; Prokop, M.; Prutyanov, V.; Puzzuoli, D.; P{\'e}rez, J.; Quintiii; Rahman, R.~I.; Raja, A.; Ramagiri, N.; Rao, A.; Raymond, R.; Redondo, R. M.-C.; Reuter, M.; Rice, J.; Riedemann, M.; Rocca, M.~L.; Rodr{\'\i}guez, D.~M.;
  RohithKarur; Rossmannek, M.; Ryu, M.; SAPV, T.; SamFerracin; Sandberg, M.; Sandesara, H.; Sapra, R.; Sargsyan, H.; Sarkar, A.; Sathaye, N.; Schmitt, B.; Schnabel, C.; Schoenfeld, Z.; Scholten, T.~L.; Schoute, E.; Schwarm, J.; Sertage, I.~F.; Setia, K.; Shammah, N.; Shi, Y.; Silva, A.; Simonetto, A.; Singstock, N.; Siraichi, Y.; Sitdikov, I.; Sivarajah, S.; Sletfjerding, M.~B.; Smolin, J.~A.; Soeken, M.; Sokolov, I.~O.; Sokolov, I.; SooluThomas; Starfish; Steenken, D.; Stypulkoski, M.; Sun, S.; Sung, K.~J.; Takahashi, H.; Takawale, T.; Tavernelli, I.; Taylor, C.; Taylour, P.; Thomas, S.; Tillet, M.; Tod, M.; Tomasik, M.; de~la Torre, E.; Trabing, K.; Treinish, M.; TrishaPe; Tulsi, D.; Turner, W.; Vaknin, Y.; Valcarce, C.~R.; Varchon, F.; Vazquez, A.~C.; Villar, V.; Vogt-Lee, D.; Vuillot, C.; Weaver, J.; Weidenfeller, J.; Wieczorek, R.; Wildstrom, J.~A.; Winston, E.; Woehr, J.~J.; Woerner, S.; Woo, R.; Wood, C.~J.; Wood, R.; Wood, S.; Wood, S.; Wootton, J.; Yeralin, D.; Yonge-Mallo, D.; Young, R.; Yu, J.;
  Zachow, C.; Zdanski, L.; Zhang, H.; Zoufal, C.; Zoufalc; a~kapila; a~matsuo; bcamorrison; brandhsn; nick bronn; brosand; chlorophyll zz; csseifms; dekel.meirom; dekelmeirom; dekool; dime10; drholmie; dtrenev; ehchen; elfrocampeador; faisaldebouni; fanizzamarco; gabrieleagl; gadial; galeinston; georgios ts; gruu; hhorii; hykavitha; jagunther; jliu45; jscott2; kanejess; klinvill; krutik2966; kurarrr; lerongil; ma5x; merav aharoni; michelle4654; ordmoj; sagar pahwa; rmoyard; saswati qiskit; scottkelso; sethmerkel; shaashwat; sternparky; strickroman; sumitpuri; tigerjack; toural; tsura crisaldo; vvilpas; welien; willhbang; yang.luh; yotamvakninibm; and {\v{C}}epulkovskis, M. 2019.
\newblock Qiskit: An Open-source Framework for Quantum Computing.

\bibitem[{Abrams and Lloyd(1999)}]{PhysRevLett.83.5162}
Abrams, D.~S.; and Lloyd, S. 1999.
\newblock Quantum Algorithm Providing Exponential Speed Increase for Finding Eigenvalues and Eigenvectors.
\newblock \emph{Phys. Rev. Lett.}, 83: 5162--5165.

\bibitem[{Baertschy and Li(2001)}]{3body}
Baertschy, M.; and Li, X. 2001.
\newblock Solution of a Three-Body Problem in Quantum Mechanics Using Sparse Linear Algebra on Parallel Computers.
\newblock In \emph{SC '01: Proceedings of the 2001 ACM/IEEE Conference on Supercomputing}, 31--31.

\bibitem[{Benedetti et~al.(2019)Benedetti, Lloyd, Sack, and Fiorentini}]{qvc_paper}
Benedetti, M.; Lloyd, E.; Sack, S.; and Fiorentini, M. 2019.
\newblock Parameterized quantum circuits as machine learning models.
\newblock \emph{Quantum Science and Technology}, 4(4): 043001.

\bibitem[{Bergholm et~al.(2018)Bergholm, Izaac, Schuld, Gogolin, Alam, Ahmed, Arrazola, Blank, Delgado, Jahangiri et~al.}]{pennylane}
Bergholm, V.; Izaac, J.; Schuld, M.; Gogolin, C.; Alam, M.~S.; Ahmed, S.; Arrazola, J.~M.; Blank, C.; Delgado, A.; Jahangiri, S.; et~al. 2018.
\newblock Pennylane: Automatic differentiation of hybrid quantum-classical computations.
\newblock \emph{arXiv preprint arXiv:1811.04968}.

\bibitem[{Biamonte et~al.(2017)Biamonte, Wittek, Pancotti, Rebentrost, Wiebe, and Lloyd}]{qml}
Biamonte, J.; Wittek, P.; Pancotti, N.; Rebentrost, P.; Wiebe, N.; and Lloyd, S. 2017.
\newblock Quantum machine learning.
\newblock \emph{Nature}, 549(7671): 195--202.

\bibitem[{Chen et~al.(2020)Chen, Yang, Qi, Chen, Ma, and Goan}]{chen2020variational}
Chen, S. Y.-C.; Yang, C.-H.~H.; Qi, J.; Chen, P.-Y.; Ma, X.; and Goan, H.-S. 2020.
\newblock Variational quantum circuits for deep reinforcement learning.
\newblock \emph{IEEE Access}, 8: 141007--141024.

\bibitem[{Deng(2012)}]{mnist}
Deng, L. 2012.
\newblock The mnist database of handwritten digit images for machine learning research.
\newblock \emph{IEEE Signal Processing Magazine}, 29(6): 141--142.

\bibitem[{Galar et~al.(2011)Galar, Fern{\'a}ndez, Barrenechea, Bustince, and Herrera}]{galar2011overview}
Galar, M.; Fern{\'a}ndez, A.; Barrenechea, E.; Bustince, H.; and Herrera, F. 2011.
\newblock An overview of ensemble methods for binary classifiers in multi-class problems: Experimental study on one-vs-one and one-vs-all schemes.
\newblock \emph{Pattern Recognition}, 44(8): 1761--1776.

\bibitem[{Gambs(2008)}]{gambs2008quantum}
Gambs, S. 2008.
\newblock Quantum classification.
\newblock arXiv:0809.0444.

\bibitem[{Grant et~al.(2018)Grant, Benedetti, Cao, Hallam, Lockhart, Stojevic, Green, and Severini}]{nature}
Grant, E.; Benedetti, M.; Cao, S.; Hallam, A.; Lockhart, J.; Stojevic, V.; Green, A.~G.; and Severini, S. 2018.
\newblock Hierarchical quantum classifiers.
\newblock \emph{npj Quantum Information}, 4(1): 1--8.

\bibitem[{Havlíček et~al.(2019)Havlíček, Córcoles, Temme, Harrow, Kandala, Chow, and Gambetta}]{Havl_ek_2019}
Havlíček, V.; Córcoles, A.~D.; Temme, K.; Harrow, A.~W.; Kandala, A.; Chow, J.~M.; and Gambetta, J.~M. 2019.
\newblock Supervised learning with quantum-enhanced feature spaces.
\newblock \emph{Nature}, 567(7747): 209–212.

\bibitem[{Hellstem(2021)}]{9425825}
Hellstem, G. 2021.
\newblock Hybrid Quantum Network for classification of finance and MNIST data.
\newblock In \emph{2021 IEEE 18th International Conference on Software Architecture Companion (ICSA-C)}, 1--4.

\bibitem[{Henderson et~al.(2020)Henderson, Shakya, Pradhan, and Cook}]{henderson2020quanvolutional}
Henderson, M.; Shakya, S.; Pradhan, S.; and Cook, T. 2020.
\newblock Quanvolutional neural networks: powering image recognition with quantum circuits.
\newblock \emph{Quantum Machine Intelligence}, 2(1): 1--9.

\bibitem[{Jurcevic et~al.(2021)Jurcevic, Javadi-Abhari, Bishop, Lauer, Bogorin, Brink, Capelluto, Günlük, Itoko, Kanazawa, Kandala, Keefe, Krsulich, Landers, Lewandowski, McClure, Nannicini, Narasgond, Nayfeh, Pritchett, Rothwell, Srinivasan, Sundaresan, Wang, Wei, Wood, Yau, Zhang, Dial, Chow, and Gambetta}]{Jurcevic_2021}
Jurcevic, P.; Javadi-Abhari, A.; Bishop, L.~S.; Lauer, I.; Bogorin, D.~F.; Brink, M.; Capelluto, L.; Günlük, O.; Itoko, T.; Kanazawa, N.; Kandala, A.; Keefe, G.~A.; Krsulich, K.; Landers, W.; Lewandowski, E.~P.; McClure, D.~T.; Nannicini, G.; Narasgond, A.; Nayfeh, H.~M.; Pritchett, E.; Rothwell, M.~B.; Srinivasan, S.; Sundaresan, N.; Wang, C.; Wei, K.~X.; Wood, C.~J.; Yau, J.-B.; Zhang, E.~J.; Dial, O.~E.; Chow, J.~M.; and Gambetta, J.~M. 2021.
\newblock Demonstration of quantum volume 64 on a superconducting quantum computing system.
\newblock \emph{Quantum Science and Technology}, 6(2): 025020.

\bibitem[{Kerenidis and Luongo(2020)}]{PhysRevA.101.062327}
Kerenidis, I.; and Luongo, A. 2020.
\newblock Classification of the MNIST data set with quantum slow feature analysis.
\newblock \emph{Phys. Rev. A}, 101: 062327.

\bibitem[{Khairy et~al.(2020)Khairy, Shaydulin, Cincio, Alexeev, and Balaprakash}]{Khairy_Shaydulin_Cincio_Alexeev_Balaprakash_2020}
Khairy, S.; Shaydulin, R.; Cincio, L.; Alexeev, Y.; and Balaprakash, P. 2020.
\newblock Learning to Optimize Variational Quantum Circuits to Solve Combinatorial Problems.
\newblock \emph{Proceedings of the AAAI Conference on Artificial Intelligence}, 34(03): 2367--2375.

\bibitem[{Kingma and Ba(2014)}]{kingma2014adam}
Kingma, D.~P.; and Ba, J. 2014.
\newblock Adam: A method for stochastic optimization.
\newblock \emph{arXiv preprint arXiv:1412.6980}.

\bibitem[{Krizhevsky, Hinton et~al.(2009)}]{cifar}
Krizhevsky, A.; Hinton, G.; et~al. 2009.
\newblock Learning multiple layers of features from tiny images.

\bibitem[{Li, Song, and Wang(2021)}]{Li_Song_Wang_2021}
Li, G.; Song, Z.; and Wang, X. 2021.
\newblock VSQL: Variational Shadow Quantum Learning for Classification.
\newblock \emph{Proceedings of the AAAI Conference on Artificial Intelligence}, 35(9): 8357--8365.

\bibitem[{Li et~al.(2021)Li, Gkoumas, Sordoni, Nie, and Melucci}]{Li_Gkoumas_Sordoni_Nie_Melucci_2021}
Li, Q.; Gkoumas, D.; Sordoni, A.; Nie, J.-Y.; and Melucci, M. 2021.
\newblock Quantum-inspired Neural Network for Conversational Emotion Recognition.
\newblock \emph{Proceedings of the AAAI Conference on Artificial Intelligence}, 35(15): 13270--13278.

\bibitem[{Liu, Hao, and Yang(2007)}]{liu2007nesting}
Liu, B.; Hao, Z.; and Yang, X. 2007.
\newblock Nesting algorithm for multi-classification problems.
\newblock \emph{Soft Computing}, 11(4): 383--389.

\bibitem[{Lockwood and Si(2020)}]{Lockwood_Si_2020}
Lockwood, O.; and Si, M. 2020.
\newblock Reinforcement Learning with Quantum Variational Circuit.
\newblock \emph{Proceedings of the AAAI Conference on Artificial Intelligence and Interactive Digital Entertainment}, 16(1): 245--251.

\bibitem[{Lu et~al.(2019)Lu, Liu, Wang, Huang, Lin, and He}]{Lu_Liu_Wang_Huang_Lin_He_2019}
Lu, C.; Liu, Q.; Wang, C.; Huang, Z.; Lin, P.; and He, L. 2019.
\newblock Molecular Property Prediction: A Multilevel Quantum Interactions Modeling Perspective.
\newblock \emph{Proceedings of the AAAI Conference on Artificial Intelligence}, 33(01): 1052--1060.

\bibitem[{Monr\`as, Sent\'{\i}s, and Wittek(2017)}]{PhysRevLett.118.190503}
Monr\`as, A.; Sent\'{\i}s, G.; and Wittek, P. 2017.
\newblock Inductive Supervised Quantum Learning.
\newblock \emph{Phys. Rev. Lett.}, 118: 190503.

\bibitem[{Nguyen and Kenyon(2018)}]{8638596}
Nguyen, N.~T.; and Kenyon, G.~T. 2018.
\newblock Image Classification Using Quantum Inference on the D-Wave 2X.
\newblock In \emph{2018 IEEE International Conference on Rebooting Computing (ICRC)}, 1--7.

\bibitem[{Preskill(2018)}]{Preskill2018quantumcomputingin}
Preskill, J. 2018.
\newblock Quantum {C}omputing in the {NISQ} era and beyond.
\newblock \emph{{Quantum}}, 2: 79.

\bibitem[{Schuld, Fingerhuth, and Petruccione(2017)}]{Schuld_2017}
Schuld, M.; Fingerhuth, M.; and Petruccione, F. 2017.
\newblock Implementing a distance-based classifier with a quantum interference circuit.
\newblock \emph{EPL (Europhysics Letters)}, 119(6): 60002.

\bibitem[{Schuld and Killoran(2019{\natexlab{a}})}]{schuld2019quantum}
Schuld, M.; and Killoran, N. 2019{\natexlab{a}}.
\newblock Quantum machine learning in feature Hilbert spaces.
\newblock \emph{Physical review letters}, 122(4): 040504.

\bibitem[{Schuld and Killoran(2019{\natexlab{b}})}]{Schuld_2019}
Schuld, M.; and Killoran, N. 2019{\natexlab{b}}.
\newblock Quantum Machine Learning in Feature Hilbert Spaces.
\newblock \emph{Physical Review Letters}, 122(4).

\bibitem[{Shor(1999)}]{shor}
Shor, P.~W. 1999.
\newblock Polynomial-time algorithms for prime factorization and discrete logarithms on a quantum computer.
\newblock \emph{SIAM review}, 41(2): 303--332.

\bibitem[{Smith et~al.(2019)Smith, Kim, Pollmann, and Knolle}]{many-body}
Smith, A.; Kim, M.~S.; Pollmann, F.; and Knolle, J. 2019.
\newblock Simulating quantum many-body dynamics on a current digital quantum computer.
\newblock \emph{npj Quantum Information}, 5(1).

\bibitem[{Tiwari and Melucci(2019)}]{Tiwari_Melucci_2019}
Tiwari, P.; and Melucci, M. 2019.
\newblock Binary Classifier Inspired by Quantum Theory.
\newblock \emph{Proceedings of the AAAI Conference on Artificial Intelligence}, 33(01): 10051--10052.

\bibitem[{Wang et~al.(2021{\natexlab{a}})Wang, Ding, Gu, Lin, Pan, Chong, and Han}]{wang2021quantumnas}
Wang, H.; Ding, Y.; Gu, J.; Lin, Y.; Pan, D.~Z.; Chong, F.~T.; and Han, S. 2021{\natexlab{a}}.
\newblock Quantumnas: Noise-adaptive search for robust quantum circuits.
\newblock \emph{arXiv preprint arXiv:2107.10845}.

\bibitem[{Wang et~al.(2021{\natexlab{b}})Wang, Ma, Hsieh, and Yung}]{wang2021quantum}
Wang, X.; Ma, Y.; Hsieh, M.-H.; and Yung, M.-H. 2021{\natexlab{b}}.
\newblock Quantum speedup in adaptive boosting of binary classification.
\newblock \emph{Science China Physics, Mechanics \& Astronomy}, 64(2): 1--10.

\bibitem[{Wilson et~al.(2018)Wilson, Otterbach, Tezak, Smith, Polloreno, Karalekas, Heidel, Alam, Crooks, and da~Silva}]{wilson2018quantum}
Wilson, C.; Otterbach, J.; Tezak, N.; Smith, R.; Polloreno, A.; Karalekas, P.~J.; Heidel, S.; Alam, M.~S.; Crooks, G.; and da~Silva, M. 2018.
\newblock Quantum kitchen sinks: An algorithm for machine learning on near-term quantum computers.
\newblock \emph{arXiv preprint arXiv:1806.08321}.

\bibitem[{Xiao, Rasul, and Vollgraf(2017)}]{xiao2017/online}
Xiao, H.; Rasul, K.; and Vollgraf, R. 2017.
\newblock Fashion-mnist: a novel image dataset for benchmarking machine learning algorithms.
\newblock \emph{arXiv preprint arXiv:1708.07747}.

\bibitem[{Yang et~al.(2020)Yang, Jiang, Zhang, and Sun}]{Yang_Jiang_Zhang_Sun_2020}
Yang, F.; Jiang, J.; Zhang, J.; and Sun, X. 2020.
\newblock Revisiting Online Quantum State Learning.
\newblock \emph{Proceedings of the AAAI Conference on Artificial Intelligence}, 34(04): 6607--6614.

\end{thebibliography}

\end{document}